# TECHNICAL REPORT:
# WHAT IS THE NEXT INNOVATION AFTER THE INTERNET OF THINGS?

Hung Cao

People in Motion Lab (Cisco Big Data Analytics)
Department of Geodesy and Geomatics Engineering
University of New Brunswick

## ABSTRACT

The world had witnessed several generations of the Internet. Starting with the Fixed Internet, then the Mobile Internet, scientists now focus on many type of research related to the "Thing" Internet (or Internet of Things). The question is "what is the next Internet generation after the Thing Internet?" This paper envisions about the Tactile Internet which could be the next Internet generation in the near future. The paper will introduce what is the tactile internet, why it could be the next future Internet, as well as the impact and its application on the future society. Furthermore, some challenges and the requirements are presented to guide further research on this near future field.

## 1. INTRODUCTION

Today's world had experienced several generation of the Internet. If the first generation of fixed Internet [1] which had virtually connected the infinite network of computer offered a chance for people collaboration and interaction between each other without regard for geographic location, the second generation of the Internet which is the Mobile Internet [2, 3] provided more flexibility and convenience to the user by combining telecommunication with the Internet. So people can connect anywhere, anytime with their mobile devices and accelerating the rapid increase in the number of Internet users recently. This is also paving a new path to the next generation of the Internet where every things and objects can be ubiquitously connected to the Internet to create the Internet of Things (IoT) [4]. One of the question is "what is the future Internet after the era of IoT?" Research scientists now start to discuss about the next vision in the future which is the Tactile Internet [5, 6]. The Tactile Internet is expected to create a potential market and a plethora of new business opportunities as well as applications that could reshape our life and economy. First coined in 2014 [5], Tactile Internet is considered as the communication method over the Internet with typical characteristic such as ultra-low latency in combination with high availability, reliability and security in order to mimic the same as human tactile reaction sense on the Internet environment. Therefore, the data lifecycle must be made within 1ms from the sensors to the actuators otherwise 'cyber-sickness' may occur [7] when users feel disoriented in an experience similar to the motion sickness sometimes suffered in the air or on the road. Developing new architectures and enabling extremely low-latency end-to-end communications to render the Tactile Internet vision realistic is one of the main motivation.

There are some review papers for the Tactile Internet can be found in [7, 8, 9]. Maier et al. (2016) [8] elaborates some commonalities, differences between the Tactile Internet, the Internet of Things and 5G vision, and survey some recent progress and enabling technologies proposed for the Tactile Internet. Some of the most strictly design challenges as well as specific solutions to enable the Tactile Internet revolution can be found in [9].

The purpose of this paper is to introduce about the Tactile Internet, its impact to our society in the near future as well as its challenges, infrastructure requirement to apply this new technology in real life. To this end, the paper is organized as follows. Section 2 outlined some typical foreseen and unforeseen effects of the Tactile Internet on the society, economic and cultures. Next, plurality of application fields are illustrated. Section 4 gives some challenges and requirement toward the Tactile Internet. Finally, in Section 5, conclusions and future work are drawn.

## 2. IMPACT OF THE TACTILE INTERNET ON THE FUTURE SOCIETY

The potential impact of the Tactile Internet is expected to bring a new dimension and method to human-to-machine, human-to-human interaction in a plurality of different society aspects including healthcare, education, energy, smart city, and culture.

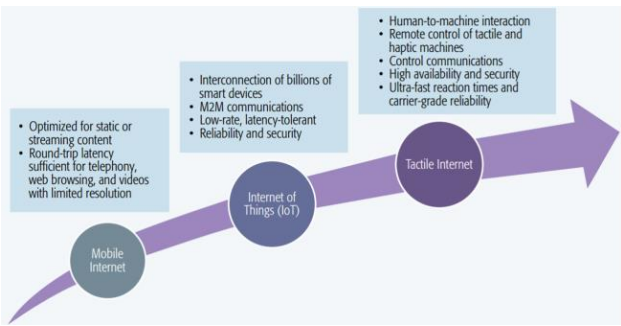

**Figure 1. The revolution leap of the Tactile Internet in the near future [8].**

Figure 1. illustrated the revolutionary leap of the Tactile Internet and listed some main driving innovations for society, economics and culture. The novel effects in which to contribute to the solution of the complex challenges faced by our society are presented in some following aspects:

- *Education:* The immediate reaction times in the Tactile Internet allows haptic overlay of the teacher and learner. As a result, novel learning experiences are driven to a new level of possibilities by the improvements in the training of specific fine-motor skills.
- *Healthcare:* By reuniting the medical experts and doctors via the Tactile Internet during remote diagnosis and treatment, the better healthcare quality could be happen. In addition, experienced surgeons' tactile sense could be combined with the high spatial precision of robot-assisted operations to cure patients in isolated environments.
- *Energy:* The Tactile Internet could be one of the foundation technologies for smart grid to improve energy efficiency and stability in electricity networks. The powerful computing capability and super agility of the Tactile Internet infrastructure enables dynamic activation and deactivation of local power generation and consumption in the decentralized electrical energy generation and distribution networks. This will potentially result in minimize the generation of unusable reactive power.
- *Smart city:* The Tactile Internet can offer some solutions for the management and operation in the Smart City. It can create of a personal spatial safety zone, or 'bubble', to assist people to interact with nearby objects. The traffic flow in the Smart City can be optimized thanks to co-operative traffic methods in the Tactile Internet. Guided autonomous driving will enable for a continuous traffic flow in which safety and energy efficiency can be significantly improved in the Smart City environment.
- *Culture:* By providing new methods in human-to-human, human-to-machine interaction, the Tactile Internet can create new behaviors, habits, perceptions not only in many aspect of culture, society, but also in the economy.

## 3. APPLICATIONS

The Tactile Internet will enhance the way of communication and lead to more realistic social interaction in various environments. In this section, some main examples are provided to show the ground-breaking potential of the Tactile Internet.

### 3.1. Virtual and Augmented Reality:

The Tactile Internet will benefit Virtual and Augmented Reality by providing the low-latency communication where several users are physically coupled via a VR or AR simulation to perform tasks together. For example, a group of artists in different country want to create a sculpture together. Currently, they are cooperating by sending the picture, design, images over the Internet to simulate a 3D prototype on the computer. They cannot directly design a real object in the real-life. But the Tactile Internet maybe the game-changing. With the robot engine, the haptic/tactile sensors, the artist can together create the specific sculpture in real-time interaction over the Tactile Internet while they are in different geographical locations.

### 3.2. Robotic and Telepresence:

Tele-presence, Tele-diagnosis, tele-surgery and tele-rehabilitation combining with the robotic technology are just some of the many potential applications of the Tactile Internet in Robotic, Healthcare service. The world had witnessed many dangerous diseases such as 'Ebola', Severe acute respiratory syndrome (SARS). People who have these kind of diseases need to be isolated. In the near future, Tactile Internet can be applied to assist the doctor and physician to diagnose, cure, surgery patients without directly contact to them. The doctor will be able to sit far from the patient's location and command the motion of a tele-robot, then receive not only audio-visual information but also critical haptic feedback.

### 3.3. Traffic:

The Tactile Internet can bring road-traffic safety and efficiency thank to its quickly computation and response characteristic. With the connected cars, the traffic light may be disappeared. Vehicles will detect a highly dynamic object such as a pedestrian, obstacle, another car by radar, lidar, sonar or video technology, and disseminate this information to neighboring vehicles within 1 ms latency. So vehicles could slide through intersections without crashing. Also, the Tactile Internet allow a pedestrian enable a personal bubble with his/her cell phone, which makes sure that no car hits the person in vicinity when crossing roads.

### 3.4. Education:

The improvement in learning experiences for students and teachers can be gained by applying Tactile Internet. Typically, this new technology can be apply in the music instrument teaching class. Teacher can stay far or close to the class but they can go over the Tactile Internet combining with Virtual/Augmented Reality to teach the students hand-by-hand or to show them how to play the music instruments.

### 3.5. Industry and Energy:

The Tactile Internet can be applied in smart grid where the latency is strictly requirements. Intelligent monitors can be achieved to optimize consumers' power supply based on information on the status of the power grid over the Tactile Internet. As a result, associated costs can be reduced. Besides, this new technology can be applied in assembly-line at the industrial zone and other applications to manufacture the mass production.

## 4. CHALLENGES AND INFRASTRUCTURE REQUIREMENTS

To achieve the envisioned Tactile Internet, there are some major problems and the architecture design requirements need to be solved. The biggest challenge is how to provide ultra-low end-to-end latency of 1ms along with the highest possible reliability for real-time response of the data lifecycle in this technology. Figure 2. presents a typical latency budget of a data lifecycle from value sensing to decisions making on the Tactile Internet environment. Besides, data security, the availability and dependability of systems that non-violate the low latency requirement due to encryption and authentication delays are the important criteria.

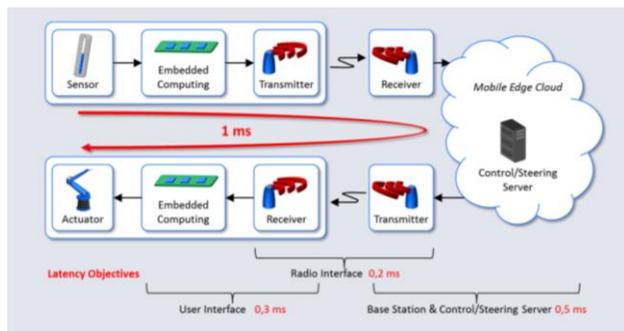

**Figure 2. The latency budget of a data life cycle in the Tactile Internet [7].**

These technical requirements of the Tactile Internet pose tremendous challenges for communications systems. The physical limit of the speed of light causing communication delay needs to be considered as well. For example, the maximum distance for a steering/control server to be placed from the point of tactile interaction by the users is 150km away (speed of light = 300km/ms) [5]. Paradigm shifts are necessary to overcome these obstacles. By distributed and decentralized service platform architecture to keep tactile applications local and close to the users, the combination of Mobile Edge-Clouds, Mini Cloud and a multi-stage hierarchy of cloud platform can be deployed as Figure 3.

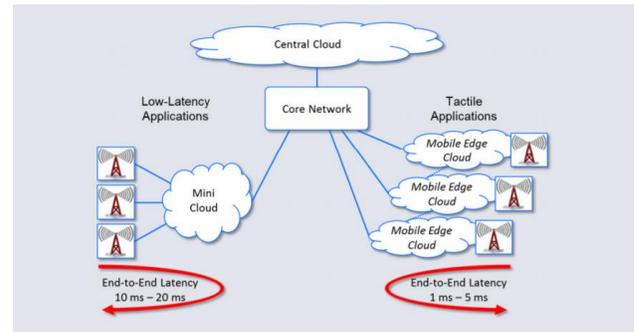

**Figure 3. The typical architecture of the Tactile Internet's infrastructure [7].**

Additionally, communications from sensors to actuators is associated with extremely demanding requirements on both hardware and software. Intelligent data compression methods combining with *"in-network processing"* such as Network Coding, Software Defined Network [10] at every node between the source and the destination can speed up the process of decision making and reduce the delay. Furthermore, new ideas and concepts to boost the requirement of stringent latency, reliability, and also capacity in both wired and wireless access network. Last but not least, edge intelligence to facilitate predictive caching as well as interpolation/extrapolation of human actions is the solution to overcome 1ms-at-speed-of-light-limit. This allow spatially decouple the active and reactive end(s) of the Tactile Internet since the tactile experience is virtually emulated on either ends. As a result, allows a much wider geographic separation between the tactile ends. It is necessary to develop novel artificial intelligence techniques based on predictive actuation for edge cloud architectures.

## 5. CONCLUSIONS

This paper discuss the next vision after the era of IoT which is the Tactile Internet. The big challenges of the Tactile Internet are extremely low-latency, reliability and high availability of the data life cycle. To achieve this, some infrastructure and technology requirements need to be performed such as decentralized architecture, in network processing, improving access networks, and applying intelligence at the edge. In the near future, the Tactile Internet is still in its fancy and can bring big impacts on education, healthcare, energy, smart city, culture and many other unforeseen of aspect of society.